\documentclass[final,3p,times]{elsarticle}

\usepackage{graphicx}
\usepackage{epstopdf}
\usepackage{color}
\usepackage{lscape}

\usepackage{amssymb}
\usepackage{amsmath}
 \usepackage{amsthm}

\journal{}

\begin{document}
\begin{frontmatter}


%
%
\title{Field-insensitive heavy fermion features and phase transition in the caged-structure quasi-skutterudite Sm$_3$Ru$_4$Ge$_{13}$}
\author{Harikrishnan S. Nair\corref{cor1}$^{1*}$, Ramesh Kumar K.$^1$, Douglas Britz$^1$, Sarit K. Ghosh$^{1,2}$, Christian Reinke$^3$, Andr\'{e} M. Strydom$^{1,4,5}$}
\address{$^1$Highly Correlated Matter Research Group, Physics Department, P. O. Box 524, University of Johannesburg, 
Auckland Park 2006, South Africa}
\address{$^2$Department of Applied Physics, Birla Institute of Technology, Mesra-835215, Ranchi, Jharkhand, India}
\address{$^3$Analytical Facility, P. O. Box 524, University of Johannesburg, Auckland Park 2006, South Africa.}
\address{$^4$Institute of Physics, Chinese Academy of Sciences, PO Box 603, Beijing 100190, China}
\address{$^5$Max Planck Institute for Chemical Physics of Solids (MPICPfS), N\"{o}thnitzerstra{\ss}e 40, 01187 Dresden, Germany}
\cortext[cor1]{h.nair.kris@gmail.com, hsnair@uj.ac.za}
\begin{abstract}
The robust field-insensitive heavy fermion features in Sm$_3$Ru$_4$Ge$_{13}$ and the magnetic phase transition at $T_N \approx$ 5~K are studied using magnetization $M(T)$, specific heat $C_p(T)$, resistivity $\rho(T)$ and thermal conductivity $\kappa_T(T)$. The average crystal structure of Sm$_3$Ru$_4$Ge$_{13}$ conforms to the cubic space group $Pm\overline{3}n$ however, signatures of subtle structural distortions are obtained from the x ray data. The magnetic susceptibility, $\chi(T)$, follows a modified Curie-Weiss law indicating the presence of crystal fields of Sm$^{3+}$ and the significance of van Vleck terms. No sign of ferromagnetism is observed in $M(H)$ of Sm$_3$Ru$_4$Ge$_{13}$ which yields only 0.025~$\mu_\mathrm{B}$/f.u.-Sm at 2~K, 7~T. The Sommerfeld coefficient, $\gamma \approx$ 220~mJ/mol-Sm K$^2$, estimated from the analysis of low temperature specific heat suggests the formation of heavy quasi particles at low temperature. Though a ln$T$ dependence of $\rho(T)$ is observed till 60~K, the resistivity behavior is accounted for by assuming a two-band model for activated behavior of charge carriers. The field scans of resistivity, $\rho(H)$, below $T_N$ display significant nonlinearity while those above the $T_N$ are more metal-like. Low values of thermal conductivity, $\kappa_T(T)$, are observed in Sm$_3$Ru$_4$Ge$_{13}$ however, displaying an anomaly at $T_N$ which signifies magnetoelastic coupling. A fairly high value of Seebeck coefficient, $S \approx$ 40~$\mu$V/K is observed at 300~K. We identify Sm$_3$Ru$_4$Ge$_{13}$ as a low charge carrier density system with unusual field-insensitive heavy fermion features very similar to the filled skutterudites.
\end{abstract}

\end{frontmatter}
\section{\label{INTRODUCTION} Introduction}
\begin{figure}[!b]
\centering
\includegraphics[scale=0.125]{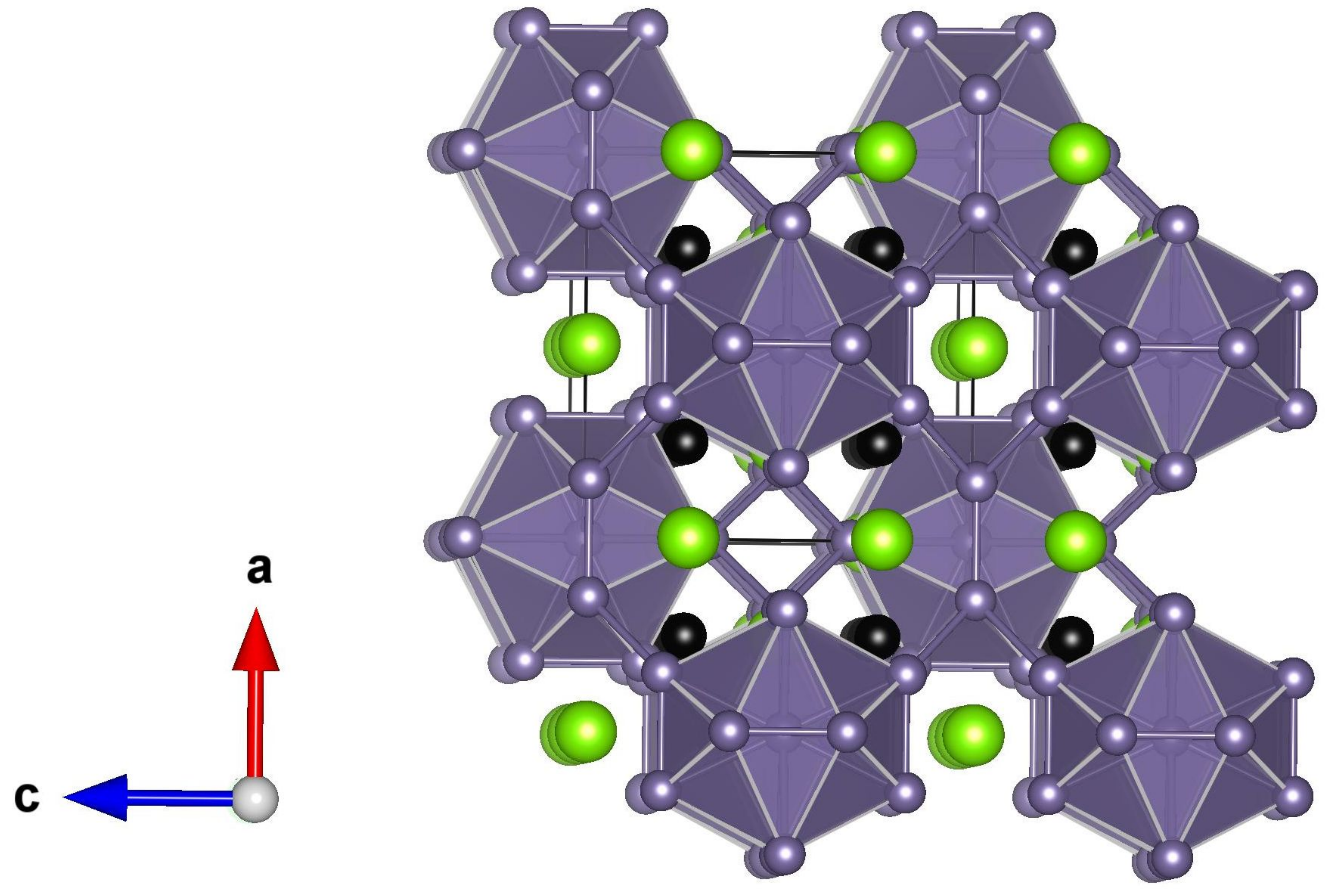}
\caption{\label{fig_str} (color online) The structural motif of  Sm$_3$Ru$_4$Ge$_{13}$ showing the Ge-polyhedra forming the caged structure. The green spheres are Sm, the black are Ru and the grey are Ge. The figure was prepared using VESTA\cite{vesta}.}
\end{figure}

\indent
Intermetallics with caged structures stimulate deep 
interest in strongly correlated electron physics 
because of the possibility of heavy fermion physics, 
superconductivity, "rattling" of atoms and thermoelectricity
\cite{uher_ss_69_2001skutterudites,sales_2003filled,nolas_arms_29_1999skutterudites,leithe_prl_91_2003ferromagnetic,matsuhira_jmmm_310_2007sm,ogita_jmmm_310_2007raman}.
The caged structure are also known for field-insensitive heavy
fermion state\cite{sanada_jpsj_74_2005exotic,yuhasz_prb_71_2005heavy} and
mixed-valence\cite{mizumaki_jpsj_76_2007mixed,yamasaki_prl_98_2007coexistence}.
For example, the filled-skutterudites of chemical formula 
$RT_4X_{12}$ ($R$ = rare earth, $T$ = transition metal 
and $X$ = P, As or Sb) display heavy fermion superconductivity, 
multipole ordering and unusual field-insensitive 
features in specific heat and in resistivity
\cite{sato2009chapter,bauer_prb_65_2002superconductivity,sanada_jpsj_74_2005exotic} 
which are also observed in the Frank-Kasper type 
caged structure intermetallics, $RT_2X_{20}$
\cite{yamada_jpsj_82_2013anomalously,isikawa_jpsj_83_2014anomalous,sakai_jpsj_81_2012low}.
Both $RT_4X_{12}$ and $RT_2X_{20}$ display puzzling 
physics especially when $R$ = Sm
\cite{higashinaka2011unusual,aoki2007specific,matsuhira2005specific,matsuhira_jmmm_310_2007sm,sanada_jpsj_74_2005exotic,takeda_physica_2003ferromagnetic}.
The field-insenstive phase transition in some of 
the Sm-based $RT_2X_{20}$ has been attributed to 
octupolar ordering\cite{higashinaka2011unusual}.
Though Sm-based heavy fermion systems are limited 
({\em e.g.,} SmPd$_3$ with $\gamma$ = 280~mJ/mol K$^2$ and
SmFe$_4$P$_{12}$ with $\gamma$ = 370~mJ/mol K$^2$), 
exotic heavy fermion state has been reported in the filled 
skutterudites SmOs$_4$Sb$_{12}$\cite{sanada_jpsj_74_2005exotic}
and SmRu$_4$P$_{12}$\cite{aoki2007specific}.
\\
\indent
The ternary intermetallic compounds denoted by 
the nominal stoichiometry $R_3T_4X_{13}$ [where $R$ is 
either a rare-earth element, an early $d$-block element 
such as Sc or Y, or an alkali-earth metal like Ca or Sr; 
$T$ stands for a Group VIII $d$-electron 
element and $X$ is In, Ge, or Sn] with $R$ = Sm 
are less-explored systems in connection with heavy 
fermion physics or field-insensitive phase transitions. 
The first report on $R_3T_4X_{13}$ phase is generally 
accredited to Remeika {\em et al,}\cite{remeika_ssc_34_1980new}.
The crystal structure of the $R_3T_4X_{13}$ series is identified by the 
cubic space group $Pm\overline{3}n$ which is common to most of
the compounds in this class, but the literature 
refers to the archetypal $R_3T_4X_{13}$ phase as either the 
Pr$_3$Rh$_4$Sn$_{13}$-structure type \cite{vandenberg_mrb_15_1980}, or the 
Yb$_3$Rh$_4$Sn$_{13}$-type\cite{hodeau_ssc_36_1980crystal}.  
This structure type allows a single crystallographic site 
for occupation by each of the $R$ and $T$ atoms, 
and two distinct sites are available to the $X$ atom. 
Following the initial discovery, $R_3T_4X_{13}$ attracted 
attention mainly due to the discovery of superconductivity 
with superconducting temperatures as high as 8 K being reported in Yb$_3$Rh$_4$Sn$_{13}$ 
and Ca$_3$Rh$_4$Sn$_{13}$\cite{remeika_ssc_34_1980new}, and 
has remained a topic of recent interest
\cite{kase_physica_471_2011,kulkarni_prb_84_2011crossover}.
In Y$_3$Ir$_4$Ge$_{13}$, peculiar structural aspect 
of the formation of atomic cages was illustrated
and its effect on thermal transport and other 
properties related to lattice vibration modes
were elucidated\cite{strydom_jpcm_19_2007thermal}. 
Charge ordering has also been suggested in this class of materials, 
for example, in Yb$_3$Ir$_4$Ge$_{13}$
\cite{strydom_physica_403_2008r3ir4ge13}.
\\
\indent
A schematic of the crystal structure of Sm$_3$Ru$_4$Ge$_{13}$ 
is shown in Fig~\ref{fig_str} highlighting 
the caged structure motifs. In the figure, 
the big green spheres are Sm, the 
small black spheres are Ru and the gray spheres 
forming the polyhedra are the 
Ge atoms. Two kinds of cages -- one centered at Ge ($2a$ position) and 
Sm ($6c$ position) -- are shown as gray cages. 
The empty Ge-cage is not 
depicted in the figure. A clear understanding of subtle 
details of distortions in the crystal structure 
is essential to understand the physics of $R_3T_4X_{13}$. 
Remeika {\em et al.,} first reported the crystal 
structure of $R_3T_4X_{13}$ in a cubic primitive 
unit cell however, also suggested 
plausible face-centered and tetragonal structures
\cite{remeika_ssc_34_1980new}.
Hodeau {\it et al.,} reported a body-centered 
unit cell for the same compound
\cite{hodeau_ssc_42_1982structural}. Modifications 
to the cubic structure with changes 
in stoichiometry were suggested\cite{eisenmann_jlcm_1986cage}. 
The first two reports on the crystal structure of the 
germanides are from Segre {\em et al.,}
\cite{segre_1981properties} and from Bruskov {\em et al.,}
\cite{bruskov_22_1986crystal} 
who proposed a disordered variant of the crystal 
structure where Ge atoms partially occupy two 
different 24$k$ sites. Positional disorder of the 
$X$ atom or the formation of random mixture of $R/T$ and 
$X$ atoms is also known
\cite{mudryk_jpcm_13_7391_2001physical,niepmann_znatur_2001structure}.
Doubling of the unit cell to a non-centrosymmetric $I4_132$ space group 
has also been observed in some of 
the stannides of $R_3T_4X_{13}$
\cite{nagoshi_jpsj_75_2006magnetic,bordet_ssc_78_1991synchrotron}.
In this paper we investigate a Sm-based 
quasi-skutterudite named
Sm$_3$Ru$_4$Ge$_{13}$ motivated to explore 
similar physics as has been
observed in $RT_4X_{12}$ or $RT_2X_{20}$ concerning 
exotic heavy fermions states or the field-insensitive 
phase transitions. In fact, through our
structural, magnetic, thermodynamic and transport 
investigations of the title
compound we observe robust field-insensitive 
heavy fermion features.
\section{\label{EXPERIMENTAL} Experimental Details}
Polycrystalline samples of 
Sm$_3$Ru$_4$Ge$_{13}$ were prepared 
by arc melting the constituent elements 
Sm ($4N$), Ru ($4N$) and Ge ($4N$) 
together, according to stoichiometric ratio, 
in an Edmund Buehler arc melting 
furnace under static atmosphere of purified argon gas. 
A Zr-getter trap was used for purifying the Ar gas. 
The once-melted sample was re-melted 4 times in 
order to ensure a homogeneous single phase. 
The weight loss of the samples
after arc melting was found to be not greater 
than 3$\%$. The melted samples, 
wrapped in Ta foils, were annealed in an evacuated 
quartz tube for 7 days at 
900~$^\circ$C. Powder X ray diffractogram 
on pulverized sample of 
Sm$_3$Ru$_4$Ge$_{13}$ was obtained using a 
Philips X'pert diffractometer using 
Cu K$\alpha$ radiation. Structural analysis of 
the powder data was performed using 
Rietveld method\cite{rietveld} implemented in 
FullProf suite of programs\cite{carvajal}.
The phase purity was probed using 
backscattered-electron imaging and X-ray 
microanalysis using a Cameca SX
100 electron microprobe (EPMA). The electron
microprobe analysis was performed on optically 
polished samples, measuring the intensities of 
the L$\alpha$ lines of the rare-earth
elements, the K$\alpha$ lines of the transition 
metals and the second-order K$\alpha$ line of 
germanium with wavelength-dispersive
spectrometers.  The probe was calibrated on 
rare-earth orthophosphates
\cite{jarosewich_newsletter,donovan_2002}
and pure metals, respectively. 
The measured X-ray intensities
were converted to weight fractions using the 
X-PHI matrix correction in
Cameca's Peak Sight software (ver. 4.0)
\cite{merlet_2002}.
Magnetic measurements were performed using a 
commercial Magnetic Property 
Measurement System and specific heat was 
measured using a commercial Physical 
Property Measurement System 
(both instruments from M/s Quantum Design).
The specific heat of a non-magnetic reference
compound Sm$_3$Ru$_4$Ge$_{13}$ was also measured. 
Electrical resistivity was measured on a 
bar-shaped sample using the ac transport 
option of the PPMS. A similar bar-shaped 
sample (roughly 7.5~mm $\times$ 2~mm $\times$ 1~mm) 
was used for thermal conductivity 
measurements using the Thermal Transport Option of 
the PPMS which measures the thermal conductivity, 
electrical resistivity and Seebeck coefficient simultaneously.
\section{\label{RESULTS} Results}
\subsection{\label{crystalstructure} Crystal structure}
\begin{figure}[!b]
\centering
\includegraphics[scale=0.125]{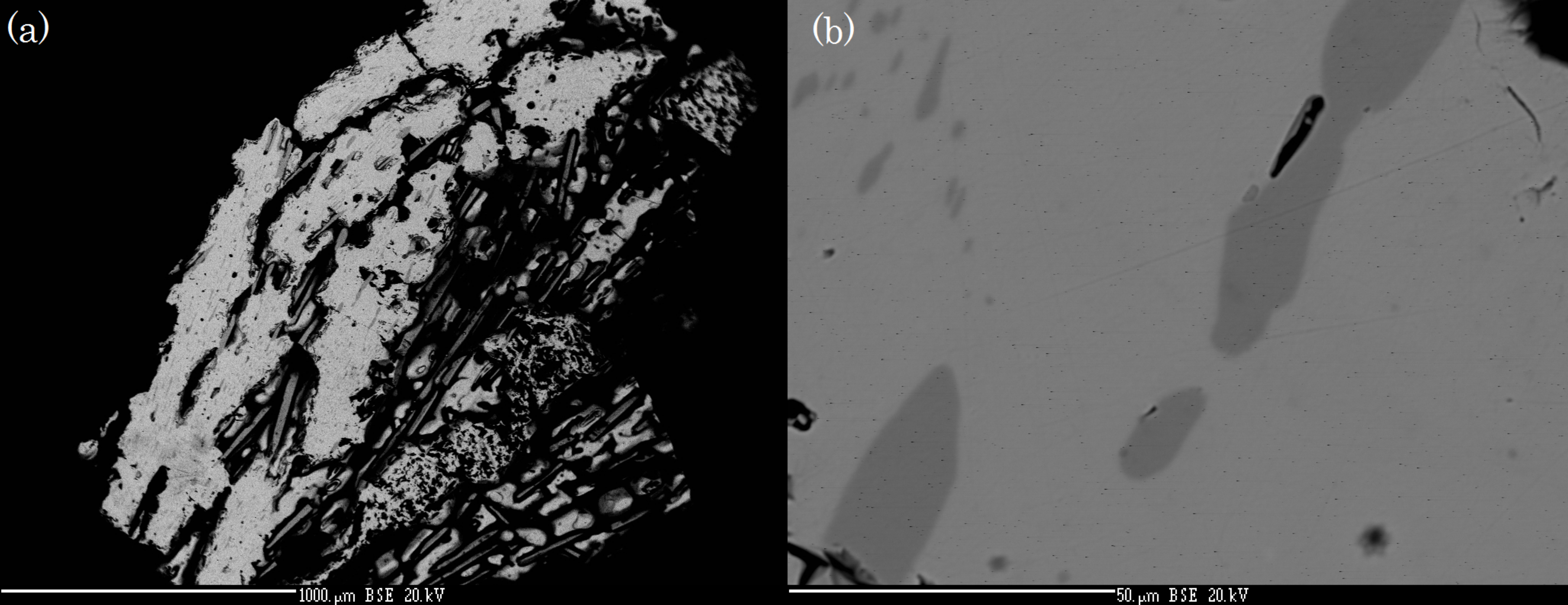}
\caption{\label{fig_epma} (color online) The metallographic image obtained from the surface of Sm$_3$Ru$_4$Ge$_{13}$ through electron microprobe analysis. Different magnifications are shown in (a) and in (b). The darker regions in (b) correspond to Ru$_2$Ge$_3$ binary phase.}
\end{figure}

\begin{figure}[!b]
\centering
\includegraphics[scale=0.33]{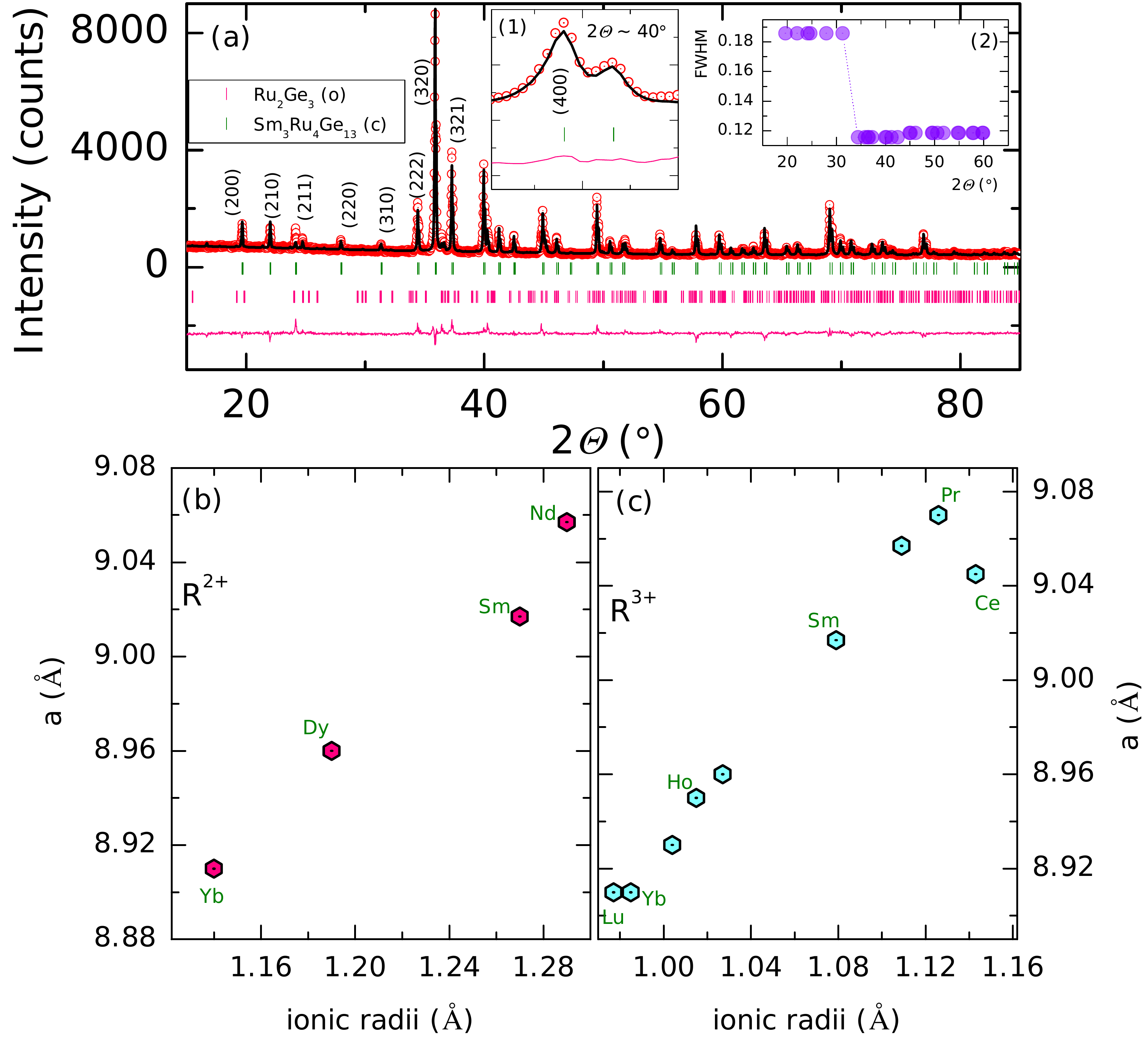}
\caption{\label{fig_xrd} (color online) (a) The x ray diffraction pattern of Sm$_3$Ru$_4$Ge$_{13}$ (red circles). The black solid line is the calculated pattern assuming $Pm\overline{3}n$ space group for Sm$_3$Ru$_4$Ge$_{13}$ and the binary phase, Ru$_2$Ge$_3$ (orthorhombic (o)). The vertical tick-marks are the allowed Bragg positions and the difference pattern is shown as a solid line. The inset (1) of (a) shows enlarged region around 40$^{\circ}$ and (2) presents a plot of FWHM vs 2$\Theta$. Plots showing the lattice parameter variation of the $R_3$RuGe$_{13}$ as a function of ionic radius of (b) $R^{2+}$ and (c) $R^{3+}$ with Sm-values incorporated from the present work. The data for other $R$'s are obtained from Ref.\cite{ghosh_prb_48_1993resistivity}.}
\end{figure}

Following the electron microprobe analysis (EPMA), 
the elemental concentration of the sample was ascertained
as Sm$_{3.54}$Ru$_{3.91}$Ge$_{11.85}$ which is
in close proximity with the nominal composition
but with a deficiency of Ge. The chemical composition
is estimated as the average from 7 different positions
on the sample surface. A photograph of the 
microstructure in two different magnifications are 
shown in Fig~\ref{fig_epma} (a) and (b). The darker 
phase seen in (b) corresponds to the binary Ru$_2$Ge$_3$
in orthorhombic $Pbcn$ space group. 
As mentioned in Section~\ref{INTRODUCTION}, 
the crystal structure of $R_3T_4X_{13}$ 
compounds is commonly described in the 
cubic space group $Pm\overline{3}n$
($\#$223). Recent crystallographic studies on 
$R_3T_4X_{13}$\cite{gumeniuk_dalton_41_20123} 
as well as early work\cite{hodeau_ssc_36_1980crystal} point towards 
subtle structural distortions and variants which may be present
in these structure types. Rietveld refinement trials 
of the x ray data on Sm$_3$Ru$_4$Ge$_{13}$ was performed first 
using the space group $Pm\overline{3}n$. Subsequently, additional
phases of Ru$_2$Ge$_3$ ($Pbcn$)
was added in the refinement cycles.
The results of the refinement are presented in Fig~\ref{fig_xrd} 
where the experimentally observed data are shown in circles, 
the calculated pattern in thick solid line 
and the allowed Bragg positions as vertical 
tick marks. The Bragg peaks corresponding to the major phase of
Sm$_3$Ru$_4$Ge$_{13}$ (88$\%$) and the minor phases Ru$_2$Ge$_3$ (12$\%$)
are differentiated using different colors in the figure.
The magnetic behaviour of Ru$_2$Ge$_3$ is described as 
temperature-independent in the temperature range 50-900~K
\cite{susz1980diffusionless,hayward2002effect}.
We conclude that this binary phase which unintentionally 
co-melted with the major phase is of no consequence in the 
analysis and findings of the results obtained on the title 
compound Sm$_3$Ru$_4$Ge$_{13}$.
In the inset (1) of panel (a), the (400) reflection from
the cubic unit cell of Sm$_3$Ru$_4$Ge$_{13}$ is shown
magnified. A comparison with the report of Gumeniuk {\em et al.,}
\cite{gumeniuk_dalton_41_20123} shows that a possible
tetragonal distortion that splits the cubic (400)
reflection in to tetragonal (440) and (004) reflections
may be present. The evolution of full-width-at-half-maximum
(FWHM) with 2$\Theta$ is shown in the inset (2) where
an abrupt change is observed close to 40 $^{\circ}$. These
observations hint at possible structural distortions
that could be present, however, requires detailed
high resolution scattering data to confirm.
The details of the structural model including the 
atomic coordinates, refined lattice parameters and goodness-of-fit are 
collected in Table~\ref{tab1}. The lattice parameters
of the binary phases are also given. The isotropic 
thermal parameters, $B_\mathrm{iso}$, were 
collected from Ref.\cite{yang2015kondo}
and the occupancies were fixed to nominal values. 
In order to ascertain whether other structural variants 
or distortions plausible for $R_3T_4X_{13}$ 
are present in Sm$_3$Ru$_4$Ge$_{13}$, 
we performed Rietveld refinement trials 
using permitted space groups 
$P4_2cm$, $P4_2mcm$, $I2_13$, $Im\overline{3}$ 
for this type of structure
\cite{gumeniuk_dalton_41_20123,hodeau_ssc_36_1980crystal}. 
The unit cell parameters,
$a$ = $b$ = 12.7497, $c$ = 9.0214 were 
used for $P4_2cm$ and $P4_2mcm$;
$a$ = 18.0578 for $I2_13$ and $Im\overline{3}$. 
It was noticed that the $P4_2cm$ space group offered a 
reasonable fit to the observed data however, the 
intensities were not completely faithfully 
accounted for. Within the limited resolution 
of our laboratory x ray data, it was not possible to 
quantify the subtle structural distortions present in 
Sm$_3$Ru$_4$Ge$_{13}$. 
\subsection{\label{heatcapacity} Heat capacity}
\begin{figure}[!t]
\centering
\includegraphics[scale=0.45]{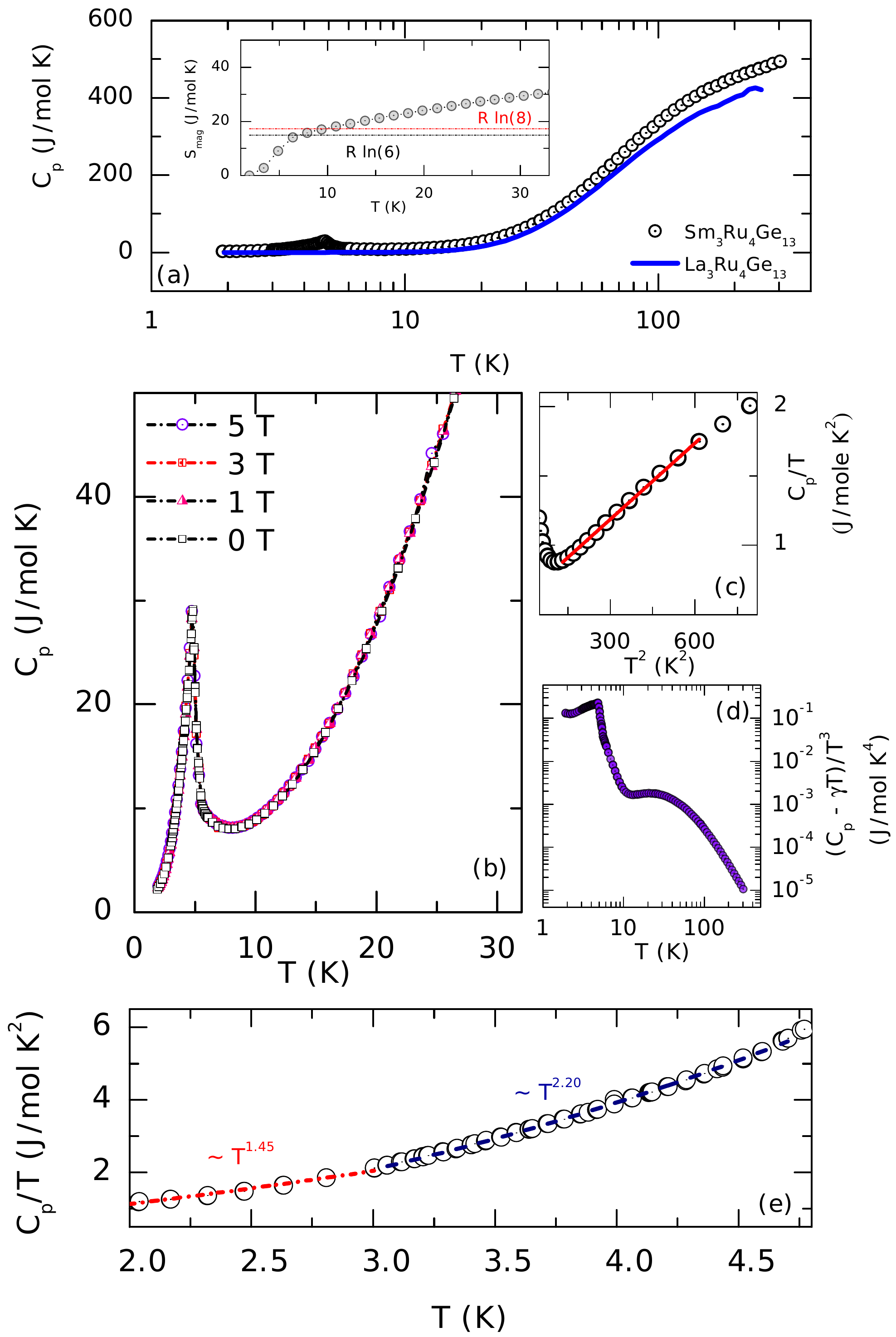}
\caption{\label{fig_cp} (color online) The specific heat $C_p$ of Sm$_3$Ru$_4$Ge$_{13}$ is presented in (a) as black open circles and that of the non-magnetic La$_3$Ru$_4$Ge$_{13}$ as a blue solid line. The inset displays the magnetic entropy, $S_\mathrm{mag}$(b) Field-dependent specific heat reveals the robustness of $C_p(T)$ up to 5~T. (c) Shows a plot of $C_p/T$ versus $T^2$ used for a linear fit (solid line) which yielded $\gamma \approx$ 220~mJ/mol-Sm K$^2$. (d) Shows a log-log plot of $(C_p(T) - \gamma T)/T^3$ versus $T$ which shows only a weak and broad feature from phonon modes centered about 30~K. (e) $C_p(T)/T$ below $T_N$ with two regions displaying different temperature-dependencies {\em viz.,} $T^{1.45}$ and $T^{2.20}$.}
\end{figure}

The experimentally measured zero-field specific 
heat of Sm$_3$Ru$_4$Ge$_{13}$ is 
presented in Fig~\ref{fig_cp} (a) as circles.  
A prominent $\lambda$-like transition is observed 
at $T_N \approx$ 5~K.  In the high-temperature 
region $T >$ 300~K, the Dulong-Petit 
value 3$nR$ $\approx$ 498~J/mol K is recovered ($n$ is 
the number of atoms in the unit cell and 
$R$ is the universal gas constant). Also shown
in (a) is the specific heat of the non-magnetic
analogue, La$_3$Ru$_4$Ge$_{13}$ as a blue solid line.
The inset of (a) shows the magnetic entropy $S_\mathrm{mag}$
estimated after subtracting the specific heat
of La$_3$Ru$_4$Ge$_{13}$ from that of Sm$_3$Ru$_4$Ge$_{13}$
The values of $S$ = $R$ln(2$J$ + 1) corresponding to 
$J$ = 5/2 and 7/2 are marked in the figure with dotted
horizontal lines.
\indent
Figure~\ref{fig_cp} (b) shows the experimental 
$C_p(T)$ measured at applied fields of 
1, 3 and 5~T. The zero-field $C_p$ is also 
re-plotted in this graph for the sake of 
completeness. It is remarkable to note that 
the specific heat is insensitive to 
changes in applied magnetic field; up to 
5~T in the present case. Such 
field-insensitive behavior is observed in Sm-based caged 
structure systems like SmTa$_2$Al$_{20}$ and SmTi$_2$Al$_{20}$
\cite{yamada_jpsj_82_2013anomalously,higashinaka2011unusual}.
The robustness of $C_p(T)$ observed under 
magnetic field for Sm$_2$Ta$_2$Al$_{20}$
was explained in the parlance of multipole 
moments in the quartet ground state of Sm ion. 
A collection of relevant thermodynamic parameters 
of some of the $R_3T_4$Ge$_{13}$ 
are presented in Table~\ref{tab2}. It can be 
readily seen that 
Sm$_3$Ru$_4$Ge$_{13}$ has a high $\gamma$ values 
among most of the Ge-based compounds.
Fig~\ref{fig_cp} (c) shows the plot of 
$C_p/T$ versus $T^2$. The linear 
region immediately above the $T_N$ was 
fitted with the expression 
$C_p(T)$ = $\gamma T$ + $\beta T^3$ 
(shown in red solid line). 
The value of the Sommerfeld 
coefficient $\gamma$ estimated from the 
fit is 220~mJ/mol-Sm K$^2$. 
A $\gamma$ value of 14~mJ/mol K$^2$ was 
obtained for the Kondo system, Yb$_3$Os$_4$Ge$_{13}$
\cite{yang2015kondo}.
A value of 280~K was extracted for $\Theta_D$ from this low temperature fit. 
For a comparison, the $\gamma$-values obtained for some of the 
Sm-containing intermetallics are: SmV$_2$Al$_{20}$ = 720~mJ/mol K$^2$ and 
SmCr$_2$Al$_{20}$ = 1000~mJ/mol K$^2$\cite{sakai_jpsj_81_2012low}.
A significantly high value of $\gamma$ for Sm$_3$Ru$_4$Ge$_{13}$ clearly 
shows that the effective electron mass has been enhanced.
\\
\indent
In Fig~\ref{fig_cp} (d), a plot of 
$(C_p(T) - \gamma T)/T^3$ versus $T$ is shown.
In the quasi-skutterudite compounds 
with strong "rattling" modes,
such a plot reveals the phonon contribution most clearly
\cite{strydom_jpcm_19_2007thermal,strydom2014superconductivity}.
In the case where strong phonon modes are present, 
the $(C_p(T) - \gamma T)/T^3$ versus $T$
or $C_p(T)/T^3$ versus $T$ plots would produce a clear anomaly
at $T_\mathrm{max} \approx \Theta_D/5$\cite{strydom_jpcm_19_2007thermal}.
Germanide quasi-skutterudites, Y$_3$Ir$_4$Ge$_{13}$ and
1: 2: 20 alloys like LaV$_2$Al$_{20}$\cite{sakai_jpsj_81_2012low}
are known to display such features.
However, in the present case of 
Sm$_3$Ru$_4$Ge$_{13}$, strong phononic
anomalies suggestive of "rattling" are not evident.  
The low temperature specific heat in the 
magnetically ordered region below the 
$T_N$ was analyzed to test
for the type of magnetic ordering. 
Antiferromagnets display $C_p(T) \sim T^2$
or $T^2 \mathrm{exp}(-\Delta/k_BT)$ where 
$\Delta$ is the magnon gap. In the panel
(e) of Fig~\ref{fig_cp} a curve-fit is 
attempted on the low temperature specific heat
below $T_N$ using the above-mentioned equation. 
However, it was found that there
exists two regions below $T_N$ with slightly 
differing temperature dependencies and using
a modified expression,  $T^n \mathrm{exp}(-\Delta/k_BT)$, 
it was found to be $n \approx$
1.45 and 2.20, repsectively. This suggests that 
the magnetic ordering in Sm$_3$Ru$_4$Ge$_{13}$
%
%
%
\subsection{\label{magnetization} Magnetization}
\begin{figure}[!t]
\centering
\includegraphics[scale=0.43]{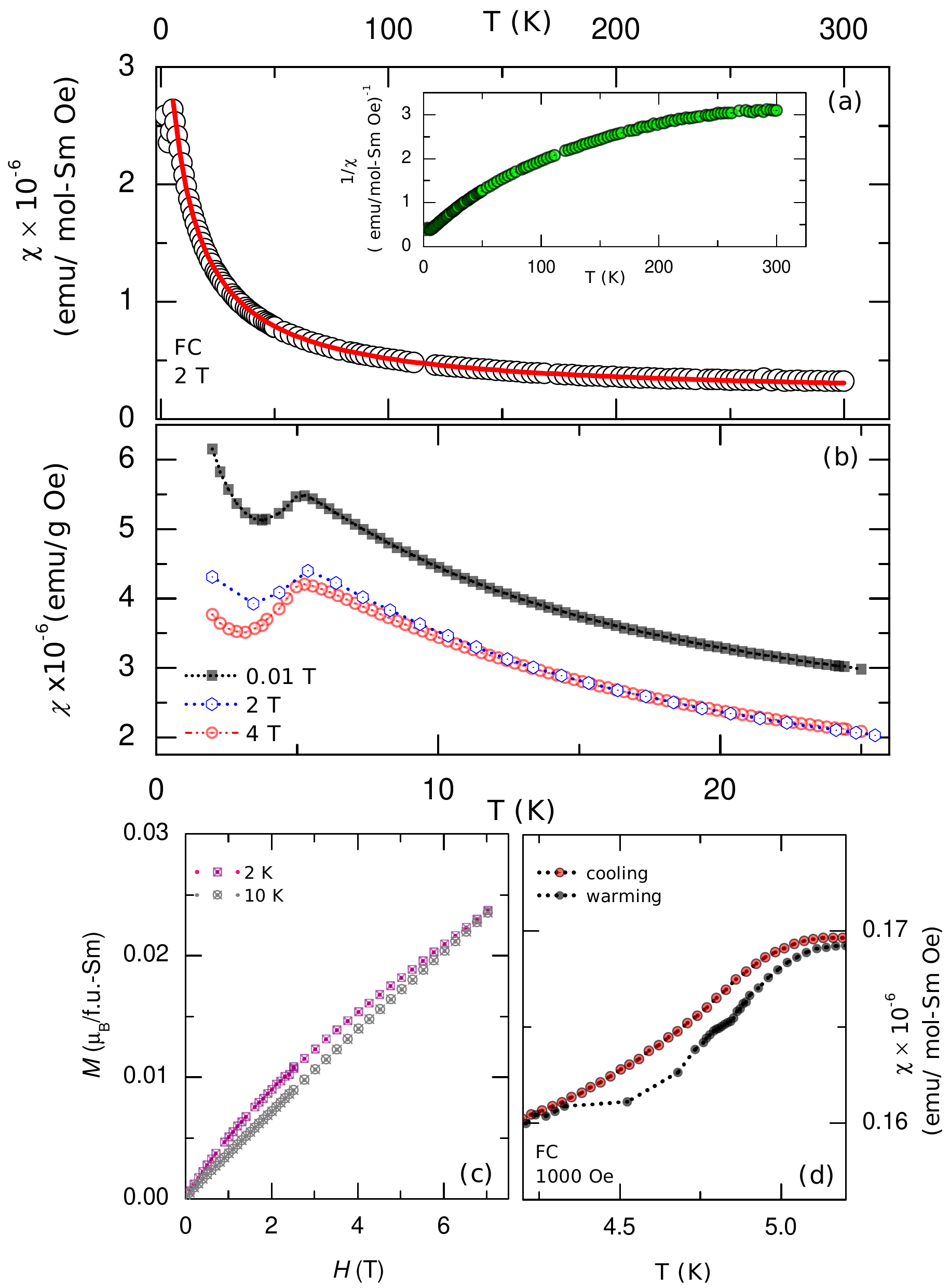}
\caption{\label{fig_mt} (color online) The magnetic response of Sm$_3$Ru$_4$Ge$_{13}$. (a) Shows the field-cooled magnetization as a function of temperature measured at 2~T applied field. An anomaly at $T_N \approx$ 5~K is observed. The solid red line in (a) is a fit assuming modified Curie-Weiss law as described in Eqn~(\ref{eqn_CW}). The inset shows the non-linear nature of 1/$\chi(T)$. (b) The $\chi(T)$ plots are various applied fields show insensitive features for higher fields. (c) Shows the field-scans of magnetization at 2~K and 10~K. (d) Shows the presence of a weak but clear thermal hysteresis around $T_N$ measured in field-cooling protocol in 0.1~T.}
\end{figure}

The magnetic response of Sm$_3$Ru$_4$Ge$_{13}$ 
is presented in Fig~\ref{fig_mt}. 
The temperature evolution of magnetic 
susceptibility in the field-cooled mode of 
measurement applying 2~T magnetic field 
is shown in Fig~\ref{fig_mt} (a) as 
black circles. The magnetic phase transition 
which was observed in the specific heat 
at $T_N \approx$ 5~K is present in the 
magnetization curve as an anomaly.
A conspicuous feature of the magnetic 
susceptibility, $\chi(T)$, of 
Sm$_3$Ru$_4$Ge$_{13}$ is the nonlinearity 
present in the whole temperature range 
of measurement which is evident in the 
1/$\chi(T)$ plot given in the inset of (a). The 
departure from Curie-Weiss behaviour may 
originate from the crystalline electric field (CEF) 
effects of the Sm$^{3+}$, taking into account 
the proximity of the $J$ = $\frac{5}{2}$ ground state and the 
first excited state with $J$ = $\frac{7}{2}$. 
The high temperature 
magnetic susceptibility of Sm compounds can 
have significant contributions from 
van Vleck terms\cite{dewijn_pr_161_1967nuclear}. 
The fact that valence fluctuations and 
crystal field effect contributions play a 
role in the magnetic susceptibility of 
Sm-based intermetallics is  shown in the 
case of SmFe$_2$Al$_{10}$ and SmRu$_2$Al$_{10}$
\cite{peratheepan_jpcm_2015}. The 1/$\chi(T)$ 
plot of SmFe$_2$Al$_{10}$ showed a broad 
plateau-like region signifying the underlying 
valence fluctuation of the Sm$^{3+}$ ions;
however, such a feature is absent in the 
present Sm-compound.
The large curvature of $\chi(T)$ in the 
case of Sm$_3$Ru$_4$Ge$_{13}$ could 
hint at significant contribution from van 
Vleck term in the total susceptibility. 
Taking all this 
into consideration, and since a simple 
Curie-Weiss formula failed to fit the 
observed susceptibility, 
fits were performed to $\chi(T)$ employing 
a modified Curie-Weiss equation:
\begin{equation}
\label{eqn_CW}
\chi(T) = \frac{N_A}{k_B} \left[ \frac{\mu^2_\mathrm{eff}}{3(T - \theta_p)} + \frac{20 \mu^2_\mathrm{B}}{7E_g} \right]
\end{equation}
In the above equation, the first term corresponds 
to Curie-Weiss contribution from the 
$J$ = $\frac{5}{2}$ ground state multiplet and 
the second term to the  temperature 
independent van Vleck contribution from the 
excited $J$ = $\frac{7}{2}$ multiplet. 
$E_g$ corresponds to the energy separation 
between the levels. However, a 
least-squares fit to the experimental $\chi(T)$ 
using Eqn~(\ref{eqn_CW}) yielded 
only 0.12(4)~$\mu_\mathrm{B}$ for 
$\mu_\mathrm{eff}$ of Sm as the only magnetic species in
Sm$_3$Ru$_4$Ge$_{13}$. This is in contrast to the 
case of SmRu$_2$Al$_{10}$ where 
$\mu_\mathrm{eff}$ = 0.85~$\mu_\mathrm{B}$ 
was obtained hinting at more or less 
localized behaviour of Sm$^{3+}$ moments.
The reduction in the effective moment from 
the Russell-Saunders value of 
0.84~$\mu_\mathrm B$ in the present case 
could arise from crystalline electric field effects. 
\\
\indent 
In Fig~\ref{fig_mt} (b), the magnetic susceptibility 
$\chi(T)$ of Sm$_3$Ru$_4$Ge$_{13}$
for $T < 25~K$ in three different applied fields 
0.01~T, 2~T and 4~T are plotted together. It can be noted
that neither the position nor the shape of the 
peak at $T_N$ is altered with the application 
of magnetic field.
Also, the curves are suggestive of the fact that 
above 2~T, a certain insensitiveness to magnetic field
sets in. The magnetic hysteresis at 2~K and 10~K 
are presented in Fig~\ref{fig_mt} (c). At 2~K, 
a slight curvature 
is observed however at 10~K, only a linear 
response is seen. A significantly reduced
value of maximum magnetic moment is observed 
in  Sm$_3$Ru$_4$Ge$_{13}$ at 2~K, 7~T. 
In Fig~\ref{fig_mt} (d), a plot of field-cooled 
magnetization curves
Sm$_3$Ru$_4$Ge$_{13}$ in cooling and warming 
cycles of measurement applying 1000~Oe 
applied field are presented. It can be 
noticed that a weak yet, distinct thermal 
hysteresis is 
present around the $T_N$. The hysteresis 
opens wide in the proximity of $T_N$ and closes 
off at about 4.2~K. It could be understood 
as a indication towards the presence of subtle 
structural modifications in Sm$_3$Ru$_4$Ge$_{13}$
\cite{gumeniuk_dalton_41_20123}. 
Moreover, a hint about close coupling 
between magnetic 
and lattice in Sm$_3$Ru$_4$Ge$_{13}$ is 
also obtained from Fig~\ref{fig_mt} (d).
\subsection{\label{resistivity} Resistivity and magnetoresistance}
\begin{figure}[!t]
\centering
\includegraphics[scale=0.36]{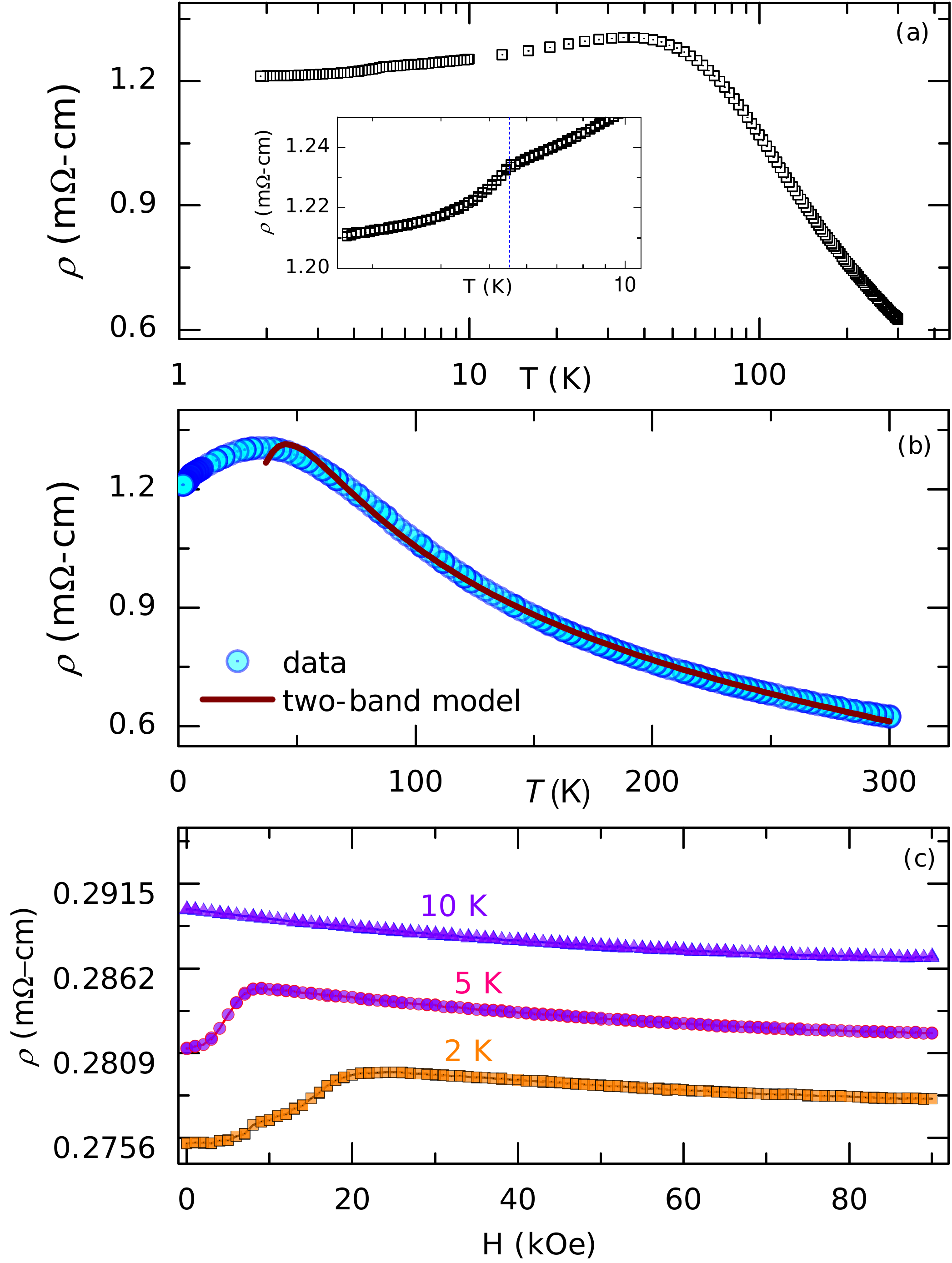}
\caption{\label{fig_rho} (color online) The electrical transport and magnetoresistance of Sm$_3$Ru$_4$Ge$_{13}$. (a) Shows the electrical resistivity as a function of temperature $\rho(T)$ plotted in semi-log axes. The phase transition at $T_N \approx$ 5~K is magnified in the inset and highlighted using a vertical line. (b) Displays the fit to $\rho(T)$ employing the two-band model as described in Eqn~(\ref{eq_twoband}). (c) The isothermal magnetoresistance of Sm$_3$Ru$_4$Ge$_{13}$ at 2, 5 and 10~K. (d) The magnetic field dependence of $\gamma$ and $A$.}
\end{figure}

\indent 
Figure~\ref{fig_rho} (a) illustrates the 
electrical resistivity, $\rho(T)$, in semi-log scale. 
A gradual increase of resistivity upon 
reduction of temperature is observed in 
Sm$_3$Ru$_4$Ge$_{13}$. At around 5~K, 
an abrupt dip is observed which 
corresponds to the magnetic phase transition 
at $T_N$ which was observed in 
magnetization as well as in specific heat 
(Fig~\ref{fig_mt} and Fig~\ref{fig_cp} respectively). 
The inset of (a) magnifies the data close 
to 5~K. The overall nature of $\rho (T)$ does 
not support that of a good metal however 
the magnitude of resistivity is comparable 
to other quasi-skutterudites in this class 
like Y$_3$Ir$_4$Ge$_{13}$\cite{strydom_jpcm_19_2007thermal}.
An activated-type of resistivity behaviour 
was observed in many $R_3T_4X_{13}$ 
systems\cite{ghosh_prb_48_1993resistivity} 
hence, an analysis 
using $\rho$ = $\rho_0\/$[1 + exp(-$\Delta$/2$T$)] 
was attempted. However, the fit 
deviated from the data at about 60~K and 
yielded a $\Delta$ value of 473.4(8)~K. 
The $\Delta$ value is comparable to that 
obtained for Y$_3$Ir$_4$Ge$_{13}$
\cite{strydom_jpcm_19_2007thermal} but 
is vastly different from systems
like CeOs$_2$Al$_{10}$\cite{lue_prb_85_2012transport}. A
curve-fit assuming $\rho \sim$ -ln (T) 
pertinent to a Kondo system was attempted
to find that the linear region extends 
over a significant range from 300~K to about 60~K.
However, magnetoresistance data (shown later) 
rules out the presence of Kondo effect. 
As a next step we analyzed the resistivity 
using a two-band model which 
describes that the prominent contribution 
to $\rho(T)$ arises from scattering 
of electrons within a broad conduction 
band and a narrow Lorentzian-shaped 
4$f$ band. In such a scenario, the electronic 
density of states can be expressed as 
$N(\epsilon_F) = W/(W^2 + P^2)$ where $W$ 
and $P$ represent the width and 
position of the $f$ band. Both $W$ and $P$ 
are temperature-dependent 
and are important in interpreting low 
temperature thermoelectric properties 
of Kondo systems. Within the two-band 
model, $\rho(T)$ is written as:
\begin{equation}
\rho(T) = \rho_0 + cT + D\frac{W}{W^2 + P^2}
\label{eq_twoband}
\end{equation}
where, $W$ = $q_f$ exp($-q_f/T$) and 
$P$ = $A$ + $B$ exp($-m/T$). $A$, $B$ and $m$ 
are constants for a given compound; $q_f$ 
is the fluctuation temperature that 
provides a measure of the quasielastic linewidth 
governing the Abrikosov-Suhl 
resonance that arises from the hybridization 
between 4$f$ and conduction band. 
The residual resistivity is indicated as 
$\rho_0$, $c$ is phonon contribution and $D$ 
represents the strength of overlap of the 
4$f$ band with the conduction band. 
In Fig~\ref{fig_rho} (b) a least-squares 
fit of the $\rho(T)$ data to Eqn~(\ref{eq_twoband}) 
is shown. It can be seen that a satisfactory 
description of resistivity data is obtained 
using the two-band model. From the fit, 
$q_f \approx$ 25~K is obtained. A similar 
value of 30~K was obtained in the case of another caged 
structure compound, CeOs$_2$Al$_{10}$
\cite{lue_prb_85_2012transport}. The two-band 
model is not valid across the phase transition 
and hence fails to explain the region 
of $\rho(T)$ below about 40~K.
\\
\indent
The isothermal magnetoresistance, $\rho$(H), of 
Sm$_3$Ru$_4$Ge$_{13}$ at 2, 5 and 10~K are presented in (c). 
The $\rho$(H) is observed to be positive. 
Interestingly, a nonlinear characteristic is displayed 
by $\rho$(H) in the temperature region below the $T_N$
with a feature-rich low-field response and a 
gradual saturation up to 90~kOe.
The $\rho(H)$ scans at 100~K and at 150~K show a
$H$-dependence very similar to the metal-like features observed at 10~K ruling
out the possibility of Kondo effect in this Sm quasi-skutterudite.
\subsection{\label{thermalconductivity} Thermal conductivity}
\begin{figure}[!t]
\centering
\includegraphics[scale=0.35]{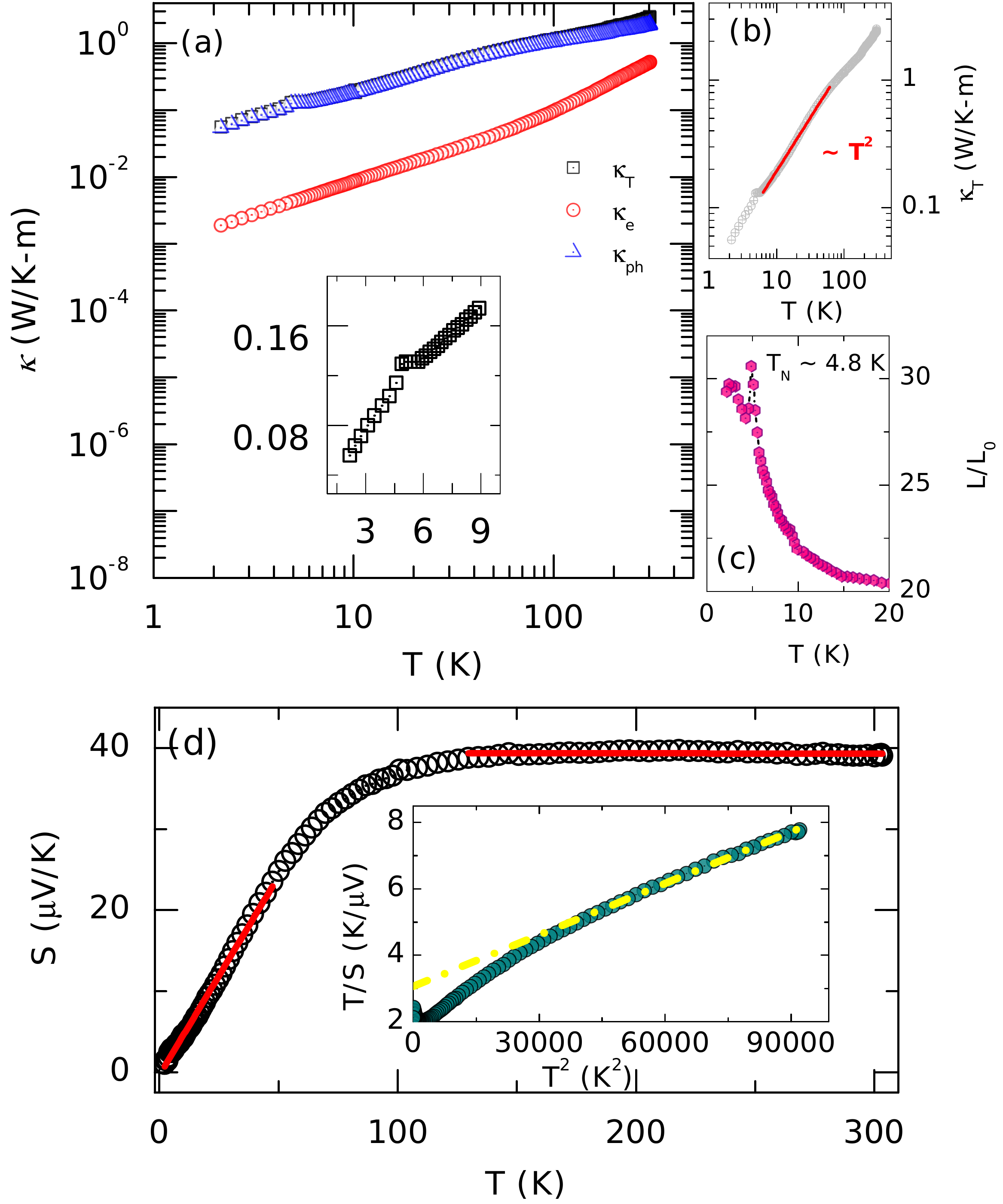}
\caption{\label{fig_tto} (color online) (a) The total thermal conductivity, $\kappa_T(T)$, of Sm$_3$Ru$_4$Ge$_{13}$ plotted along with the electronic contribution, $\kappa_e(T)$ and the phonon contribution, $\kappa_{ph}(T)$. The inset shows the magnified region around 5~K where an anomaly is identified which matches with that found in the magnetization and specific heat. (b) Shows a log-log plot of $\kappa_T(T)$ which exhibits $\sim T^2$ dependence in the region above the transition. (c) The ratio, $L/L_0$ reflecting the anomaly at $T_N$. (d) Shows the Seebeck coefficient, $S$, with a fairly large value of $\approx$ 40~$\mu$V/K at 300~K. The inset shows the plot of $T/S$ versus $T^2$ with a linear fit shown in dash-dotted line.}
\end{figure}

\indent
The total thermal conductivity $\kappa_T(T)$, 
electrical resistivity $\rho(T)$ and 
thermopower $S(T)$ were measured simultaneously 
using the thermal transport option 
of the PPMS. The temperature dependence of 
total thermal conductivity, $\kappa_T(T)$, 
of Sm$_3$Ru$_4$Ge$_{13}$ is presented in 
Fig~\ref{fig_tto} (a) as black open squares. 
The $\kappa_T(T)$ reduces by an order of 
magnitude while Sm$_3$Ru$_4$Ge$_{13}$ is cooled from 300~K to 2~K. 
The electronic contribution to the thermal conductivity, $\kappa_e(T)$ was 
extracted following the Wiedemann-Franz relation\cite{kittel_book}.
\begin{equation}
\kappa_e(T) = \frac{L_0T}{\rho(T)}
\label{eqn_kappa}
\end{equation}
In this equation, the Lorenz number is 
$L_0$ = $\left(\frac{\pi k^2_B}{e \sqrt{3}}\right)^2$ 
= 2.54 $\times$ 10$^{-8}$ W$\Omega K^{-2}$ and 
$\rho(T)$ is the electrical resistivity. 
In Fig~\ref{fig_tto} (a), $\kappa_e(T)$ is 
represented as blue open circles. The phonon 
contribution $\kappa_{ph}(T)$ to the total 
thermal conductivity was estimated using 
the expression: $\kappa_T(T)$ =  $\kappa_e(T)$ + $\kappa_{ph}(T)$. The 
$\kappa_e(T)$ estimated by Eqn~(\ref{eqn_kappa}) 
is about an order of magnitude lower than the total thermal 
conductivity. It is clear from Fig~\ref{fig_tto} 
that the thermal conductivity is dominated 
by phonons. The anomaly at $T_N$ which was 
observed in specific heat, magnetization and 
resistivity is clearly seen in thermal 
conductivity also. The region of anomaly is 
magnified in the inset of Fig~\ref{fig_tto} (a). 
\\
\indent
The total thermal conductivity, 
$\kappa_T(T)$ is plotted in a log-log scale in 
Fig~\ref{fig_rho} (b) where a relatively extended 
region of $T^2$-dependence 
is observed above the $T_N$. This feature supports 
the fact that the thermal 
conductivity is dominated by phononic contribution. 
In Fig~\ref{fig_tto} (c), a plot of normalized 
Lorenz number, $L/L_0$ is presented. Here, 
$L$ = $\left(\frac{\kappa \rho}{T}\right)$ 
where $\kappa$ and $\rho$ are the measured 
thermal conductivity and resistivity 
respectively. The anomaly observed in 
$\kappa_T(T)$ at $T_N$ is enhanced in this 
plot and signifies that additional latent 
heat is involved in the magnetic phase 
transition. However, it must be noted that 
a thermal hysteresis was observed in 
magnetization (see Fig~\ref{fig_mt} (d)) 
pointing towards the involvement 
of lattice degrees of freedom concurrent 
with the magnetic part.
The experimentally measured thermopower, 
$S(T)$, is presented in the panel
(d) of Fig~\ref{fig_tto}. At 300~K, 
$S(T)$ reaches a value of 39~$\mu$V/K. 
This value may be compared with the values 
for pure metals, 1 -- 10~$\mu$V/K, 
and with that of semiconductors, 10$^2$ -- 10$^3$~$\mu$V/K. 
The positive sign of the thermopower 
suggests that the conduction 
mechanism is hole dominated. 
The behavior of thermopower is qualitatively 
and quantitatively very similar to 
that of Y$_3$Ir$_4$Ge$_{13}$, which too, 
displayed two distinct linear regions. 
The scenario described by the two-band model 
applies to the thermopower also since the 
Seebeck coefficient is proportional to the 
energy-derivative of the density of states 
at the Fermi level. However, a dome-shaped 
feature in thermopower is not observed in 
Sm$_3$Ru$_4$Ge$_{13}$ which is 
normally observed in many compounds with 
intermediate valence states and 
CEF effects. In the inset of (d),
a plot of thermopower is presented as
$T/S$ versus $T^2$. A large region of linearity
is observed in the high temperature region. The
dash-dotted line is a linear fit to the high temperature
region above 200~K. Within the Boltzmann theory
and relaxation time approximation for an isotropic
system, the thermopower can be approximated as 
$S$ = $AT$/[$B^2$ + $T^2$] where $A$ = $\frac{2(\epsilon_0 - \epsilon_F)}{|e|}$
and $B^2$ = (3/$\pi^2 k^2_B$)(($\epsilon_0 - \epsilon_F$)$^2$ + $\Gamma^2$)
\cite{gottwick1985transport}.
From the plot of $T/S$ versus $T^2$, the constants
$A$ and $B$ can be determined. This leads to an
estimate of ($\epsilon_0 - \epsilon_F$) = 0.07~meV and
$\Gamma$ = 0.12~meV.
\section{\label{DISCUSSION} Discussion}
The present set of data on the magnetic, 
thermodynamic and transport
properties of the quasi-skutterudite, Sm$_3$Ru$_4$Ge$_{13}$,
reveal a magnetic transition at $T_N\approx$ 5~K
implying strong correlation between all 
the measured physical 
properties. The caged structure of the present 
composition is known to be conducive for 
atomic "rattling" phenomena known in this 
class of materials
\cite{strydom_jpcm_19_2007thermal,strydom_physica_403_2008r3ir4ge13}.
From the structural analysis of x ray data, 
it is concluded that the crystal 
structure conforms to the cubic space group 
$Pm\overline{3}n$ which is commonly found
in $R_3T_4X_{13}$ compounds. However, indications 
of subtle structural distortions inherent 
to this type of materials are present in the x ray data
\cite{hodeau_ssc_36_1980crystal,hodeau_ssc_42_1982structural,gumeniuk_dalton_41_20123}. 
The lattice parameters of Sm$_3$Ru$_4$Ge$_{13}$ 
estimated from the
x ray data is plotted along with the lattice 
parameters for $R_3$Ru$_4$Ge$_{13}$
taken from Ref.\cite{ghosh_prb_48_1993resistivity} 
in panel (c) of Fig~\ref{fig_xrd}. 
No sign of a structural anomaly is observed 
for the Sm compound
from such a plot. However, the superstructure-like 
peaks observed in the
x ray pattern is suggestive of subtle structural 
distortions that may be present
in Sm$_3$Ru$_4$Ge$_{13}$. 
\\
\indent
The magnetic susceptibility of Sm$_3$Ru$_4$Ge$_{13}$ 
shows indications of crystal electric field 
effects arising from the close-lying crystal 
field levels of Sm$^{3+}$ ions. The $\chi(T)$ 
behavior is qualitatively different from other 
Sm-based systems like SmFe$_2$Al$_{10}$ 
where strong valence fluctuations have been 
reported\cite{peratheepan_jpcm_2015}.
However a simple Curie-Weiss description was 
not suitable for Sm$_3$Ru$_4$Ge$_{13}$
and van Vleck terms in the total magnetic 
susceptibility were required to fit the data.
The typical response of $\chi(T)$ of compounds 
that displays valence fluctuation show 
a profound minimum which is due to the 
instability of $4f$ shell and strong
hybridization between the $4f$ electron and 
conduction electron states\cite{falkowski2015cooperative}.
The magnetic susceptibility of 
Sm$_3$Ru$_4$Ge$_{13}$ show a very
weak variation of magnitude starting from 
300~K toward 2~K. These features
rule out the possibility of valence 
fluctuation though, mixed valence of Sm
may be present. However, the field-insensitive 
features in specific heat and
the very low magnetic moment values 
attained in magnetization measurements
at low temperature point toward the possibility 
of octupolar ordering mechanisms
operating in this Sm-based 3:4:13 alloy 
similar to the case of filled skutterudites or
Frank-Kasper cages.
\\
\indent
The specific heat of Sm$_3$Ru$_4$Ge$_{13}$ displays 
strong robustness of the phase transition at 
$T_N \approx$ 5~K in applied magnetic 
fields up to 5~T. This feature is very 
similar to the observation in 
other Sm-based caged-structures, 
SmTa$_2$Al$_{20}$ and SmTi$_2$Al$_{20}$
\cite{yamada_jpsj_82_2013anomalously,higashinaka2011unusual}.
A curve-fit to the total observed specific 
heat could be advanced through a model 
description incorporating 
a Debye term, three Einstein terms and a 
linear term in specific heat. The magnetic entropy
estimated by subtracting the phonon and 
electronic part from the total specific heat
indicates the crystal field effects of Sm. 
Though the Einstein terms were required to describe the
phononic part, no indication of "rattling" is 
obtained from the plot of
($C_p - \gamma T$)/$T^3$ versus $T$. Moreover, 
a simple description of specific heat
following $C_p(T) \sim T^2$ does not hold in 
the ordered region below the $T_N$ suggesting
that Sm$_3$Ru$_4$Ge$_{13}$ may not be a simple 
antiferromagnet. The low temperature
estimate of Sommerfeld coefficient suggests 
the formation of heavy quasi particles.
A comparison of Sommerfeld coefficients and 
Debye temperatures of various Ge-based
3:4:13 alloys is presented in Table~\ref{tab2}.
\\
\indent
Though the electrical resistivity of 
Sm$_3$Ru$_4$Ge$_{13}$ exhibited 
a ln$T$ behaviour in a wide temperature above $\approx$ 60~K 
indicating possible Kondo physics, the field-dependence of 
resistivity, $\rho(H)$, at different 
temperatures rules out such a scenario.
At temperatures below $T_N$, the $\rho(H)$
is nonlinear and shows an initial 
increase in low fields which, then decreases
as the applied field is increased. However, 
above 10~K, the $\rho(H)$ shows
a monotonic decrease with applied field which 
is more metal-like. Even at 100~K and 150~K,
where the ln$T$ behavior was observed, a 
monotonic variation of $\rho(H)$
was observed (data not presented). 
Interestingly, the field-insensitive 
features observed in specific heat
was also reflected in the field-dependence 
of the Sommerfeld coefficient, $\gamma$,
and the coefficient $A$. An unusual field-
insensitiveness of $\gamma$ and $A$ up to 
5~T was clear from Fig~\ref{fig_rho} (d). 
Similar to many other Sm-based caged systems,
this field-insensitiveness might have its 
origin in octupolar ordering
\cite{aoki2007specific,higashinaka2011unusual}.
The phase transition observed in magnetization,
specific heat and resistivity at
$T_N \approx$ 5~K is also present in the 
thermal conductivity data. This might
serve as an indication for the coupling between 
all degrees of freedom in this material.
\\
\indent
The Sm-containing filled skutterudites Sm$T_4X_{12}$
($T$ = transition metal, $X$ = Ge, P, Sb) are
compounds that can be readily compared with
the present Sm-based quasi skutterudite. SmFe$_4$P$_{12}$
show Kondo behaviour\cite{takeda2008metamagnetism} 
as evidenced through the relaxation
mechanisms through $^{31}$P-NMR and $\mu$SR experiments
\cite{hachitani200631p}. The crystalline fields of Sm$^{3+}$
showed their presence as a ground state and an excited state
at 70~K and a ferromagnetic ground state was inferred to
exist below 1.6~K. Another experimental investigation
of the same compound using de Haas-van Alphen effect,
magnetoresistance and Hall effect revealed anomalous
large anisotropy in cyclotron effective mass despite
a near-spherical Fermi surface\cite{kikuchi2008anomalous}.
In this Sm-compound also, the mass enhancement was
found to be robust against applied magnetic field.
In a specific heat study combining the filled skutterudites
SmFe$_4$P$_{12}$, SmRu$_4$P$_{12}$ and SmOs$_4$P$_{12}$,
signatures of orbital and antiferromagnetic orderings
were obtained for the Ru-based compound\cite{matsuhira2005specific}.
The Pt-based filled skutterudite SmPt$_4$Ge$_{12}$
has Sm in a temperature-independent intermediate valence
state while signatures of heavy fermion and
Kondo effects are seen\cite{gumeniuk2010high}.
These reports indicate that Sm-containing skutterudites
and quasi-skutterudites are interesting compounds
where anomalous behaviour originating from rare earth
magnetism, anisotropy and dipolar interactions are observed.
\section{\label{conclusions} Conclusions}
Strong field-insensitive specific heat and heavy fermion features are observed in the quasi-skutterudite Sm$_3$Ru$_4$Ge$_{13}$. The unusual field-insensitiveness  is also present in $\gamma$ as well. Very low values of effective magnetic moment and maximum moment in field-scans of magnetization point toward possible octupolar ordering mechanisms being operative. The magnetoresistance below the $T_N$ display nonlinear characteristics but behave metal-like above the $T_N$. The features observed in the present Sm based 3:4:13 alloys bear strong resemblances to the unusual heavy fermion and field-insensitive features observed in Sm-based filled skutterudites and Frank-Kasper cages.
\section*{Acknowledgements}
HSN and RKK acknowledge FRC/URC for a postdoctoral fellowship. AMS thanks the SA-NRF (93549) and the FRC/URC of UJ for financial assistance. 
%
%
%

%
%
%
\clearpage\newpage
\begin{table}[!t]
\caption{\label{tab1} The atomic positions in the $Pm\overline{3}n$ model used to refine the x ray data of Sm$_3$Ru$_4$Ge$_{13}$. The refined lattice parameter is $a$~({\AA}) = 9.0171(2). The reliability factors of the fit are $\chi^2$ = 4.5, $R_p$ = 6.1 and $R_{wp}$ = 8.8. Wyck. stands for Wyckoff position and $B_\mathrm{iso}$ are the isotropic thermal parameters taken from \cite{yang2015kondo}. The occupancies were fixed to nominal values.}
\setlength{\tabcolsep}{9pt}
\begin{tabular}{llllll} \hline\hline
Atom & Wyck. & $x$  & $y$  & $z$  & $B_\mathrm{iso}$ \\ \hline\hline
Sm  &  $6c$ & 0.25 & 0 & 0.5 & 0.001\\
Ru & $8e$ & 0.25 & 0.25 & 0.25 & 0.001\\
Ge(1) & $2a$ & 0 & 0 & 0 & 0.001\\
Ge(2) & $24k$ & 0 & 0.1519 & 0.3149 & 0.010\\ \hline\hline
\end{tabular}\\
\begin{tabular}{lllll}
Binary             & S.G.                  &   $a$       &  $b$    &   $c$     \\ \hline\hline
Ru$_2$Ge$_3$ (o)     & $Pbcn$                & 11.4349      &  9.2423  & 5.7200  \\ \hline
\end{tabular}
\end{table}
\clearpage\newpage
\begin{table}[!b]
\caption{\label{tab2} The Sommerfeld coefficients ($\gamma$) and Debye temperatures ($\Theta_D$) of a collection of Ge-based $R_3T_4X_{13}$ quasi-skutterudites. The abbreviations are M: metal, HF: heavy fermion, SC: superconductor, sm: semi-metal, K: Kondo system.}
\setlength{\tabcolsep}{2pt}
\begin{tabular}{lllllc} \hline\hline
$R_3T_4$Ge$_{13}$          & Type             & $\gamma$                      & $\Theta_D$  & & Ref.   \\ \hline\hline
Ce$_3$Ru$_4$Ge$_{13}$  & HF & 111 mJ/mol K$^{2}$  & 187~K          &          & \cite{ghosh_prb_52_7267_1995crystal} \\
Yb$_3$Co$_{4.3}$Ge$_{12.7}$  & SC & 2.3 mJ/mol K$^{2}$  & 207~K          &          & \cite{mudryk_jpcm_13_7391_2001physical} \\
Yb$_3$Pt$_4$Ge$_{13}$  &  & $<$50 mJ/mol K$^{2}$  & 310~K          &          & \cite{gumeniuk_dalton_41_20123} \\
Yb$_3$Rh$_4$Ge$_{13}$  &    & 25 mJ/mol Y K$^{2}$  & --          &          & \cite{strydom2014superconductivity} \\
Yb$_3$Os$_4$Ge$_{13}$  & K & 14 mJ/mol Y K$^{2}$  & 160~K          &          & \cite{yang2015kondo} \\
Y$_3$Co$_4$Ge$_{13}$  & M & 5.7 mJ/mol Y K$^{2}$  & 354~K          &          & \cite{ghosh_prb_52_7267_1995crystal} \\
Y$_3$Ir$_4$Ge$_{13}$  & sm & 4.3 mJ/mol K$^{2}$  & 129~K          &          & \cite{strydom_jpcm_19_2007thermal} \\
Lu$_3$Rh$_4$Ge$_{13}$  &    & 5 mJ/mol Y K$^{2}$  & --          &          & \cite{strydom2014superconductivity} \\
Sm$_3$Ru$_4$Ge$_{13}$  & HF,K & 220 mJ/mol Sm K$^{2}$  & 280~K          &          & [This work] \\ \hline\hline
\end{tabular}
\end{table}
%
%
%
\clearpage\newpage
\clearpage\newpage
\clearpage\newpage
%
%
%
%
\clearpage\newpage
\clearpage\newpage
\clearpage\newpage
\clearpage\newpage
\end{document}